\documentclass[sigconf, screen=true, review=false, printacmref=false, printccs=false, printfolios=false]{acmart}
\setcopyright{none}

\usepackage{soul}
\usepackage{comment}
\usepackage{algorithm}
\usepackage{algorithmicx}
\usepackage{algpseudocode}
\algblockdefx[parallel]{ParFor}{EndPar}[1][]{$\textbf{parallel for}$ #1 $\textbf{do}$}{$\textbf{end parallel}$}
\algrenewcommand{\alglinenumber}[1]{\fontsize{6.5}{7}\selectfont#1}
\algtext*{EndPar}

\algblockdefx[parallel]{parfor}{endpar}[1][]{$\textbf{parallel for}$ #1 $\textbf{do}$}{$\textbf{end parallel}$}
\algrenewcommand{\alglinenumber}[1]{\scriptsize#1:}

\algblockdefx[parallel]{parallelfor}{parallelend}
[1][]{\textbf{parallel for} #1}
{\textbf{end parallel}}

\usepackage{nicefrac}

\algnotext{EndFor}
\algnotext{EndIf}
\algnotext{EndProcedure}

\usepackage{enumitem}

\usepackage{color, colortbl}

\usepackage{cleveref}

\DeclareMathAlphabet\mathbfcal{OMS}{cmsy}{b}{n}

\setcopyright{none}
\usepackage{booktabs} 
\usepackage{epstopdf}
\usepackage{comment}
\usepackage{tabularx}
\usepackage{subfigure}
\usepackage{paralist}
\usepackage{balance}

\usepackage{mathtools}
\usepackage{pifont} 
\usepackage{color}
\usepackage{xcolor}
\usepackage{array}
\newcolumntype{P}[1]{>{\centering\arraybackslash}p{#1}}
\newcolumntype{M}[1]{>{\centering\arraybackslash}m{#1}}
\newcolumntype{R}[1]{>{\arraybackslash}m{#1}}

\definecolor{orange}{rgb}{1,0.5,0}
\definecolor{graynode}{RGB}{20,20,20}
\definecolor{crimsonred}{RGB}{220,20,60}
\definecolor{darkgraynode}{gray}{0.5}
\definecolor{lightgraynode}{gray}{0.8}

\definecolor{dkgreen}{rgb}{0,0.6,0}
\definecolor{gray}{rgb}{0.5,0.5,0.5}
\definecolor{lightred}{rgb}{0.93,0.93,0.93}
\definecolor{lightred}{rgb}{0.83,0.83,0.83} 

\definecolor{lightgraydark}{rgb}{0.6,0.6,0.6}

\definecolor{googleblue}{RGB}{66,133,244}
\definecolor{googlered}{RGB}{219,68,55}
\definecolor{googlegreen}{RGB}{15,157,88}
\definecolor{insightpurple}{RGB}{15,157,88}

\definecolor{lightblue}{rgb}{0.5,0.90,1.0}
\definecolor{lightgreen}{rgb}{0.5,0.92,0.5}
\definecolor{lightyellow}{rgb}{1,0.90,0.40}

\definecolor{verylightgreen}{RGB}	{204,255,204}
\definecolor{verylightred}{RGB}		{255,204,204}
\definecolor{verylightyellow}{RGB}		{255,255,204}

\definecolor{lightgreen}{RGB}	{204,255,204}
\definecolor{lightyellow}{RGB}		{255,255,204}

\definecolor{plotblue}{RGB}	{30,144,255}
\definecolor{plotgreen}{RGB}	{50,205,50}
\definecolor{plotred}{RGB}	{220,20,60}

\definecolor{myyellow}{RGB}{255,255,204}
\definecolor{myred}{RGB}{255,204,204}
\definecolor{lightblue}{RGB}{0,200,255}
\definecolor{mygreen}{RGB}{80,220,80}

\definecolor{gray}{RGB}{20,20,20}
\definecolor{greencm}{RGB}{0,153,0}

\definecolor{theblue}{RGB}{0,0,180}

\definecolor{matlabgreen}{rgb}{0,0.6,0}
\definecolor{matlabgray}{rgb}{0.5,0.5,0.5}
\definecolor{matlabmauve}{rgb}{0.58,0,0.82}

\usepackage{rotate}
\usepackage{capt-of}
\usepackage{setspace}
\usepackage{mathtools}
\usepackage{pifont}

\definecolor{gray}{RGB}{20,20,20}
\definecolor{gray}{RGB}{0.7,0.7,0.7}
\definecolor{greencm}{RGB}{0,153,0}

\usepackage{tabularx}
\usepackage{booktabs}

\definecolor{thedarkblue}{RGB}{0,0,120} 
\definecolor{mydarkblue}{rgb}{0,0.08,0.45} 
\definecolor{darkblue}{rgb}{0,0.08,180}
\colorlet{TufteRed}{red!80!black}

\definecolor{theblue}{RGB}{0,0,180}
\colorlet{thered}{TufteRed}

\usepackage{hyperref}
\hypersetup{
colorlinks=true,
linkcolor=mydarkblue,
citecolor=mydarkblue,
filecolor=mydarkblue,
urlcolor=mydarkblue}

\usepackage{microtype}
\usepackage{balance}

\usepackage{amsmath,amsthm}

\newcommand{\incomplete}[1]{{\textcolor{red}{#1}}}
\renewcommand{\incomplete}[1]{\ignorespaces}

\newcommand{\todo}[1]{}

\newcommand{\eat}[1]{\ignorespaces}
\newcommand{\journal}[1]{\ignorespaces}

\usepackage{tikz}
\usepackage{verbatim}
\usetikzlibrary{arrows}
\usetikzlibrary{shapes,snakes}
\usetikzlibrary{decorations.pathmorphing} 
\usetikzlibrary{fit}					
\usetikzlibrary{backgrounds}	

\usepackage{ragged2e}
\usepackage{multirow}
\usepackage{microtype}
\usepackage{balance}
\usepackage{setspace}

\graphicspath{{./}{./graphics/}}
\newcolumntype{H}{>{\setbox0=\hbox\bgroup}c<{\egroup}@{}}

\newcolumntype{R}[1]{>{\RaggedLeft\arraybackslash}} 
\newcolumntype{L}[1]{>{\RaggedRight\arraybackslash}}

\usepackage{amsmath,amsthm}
\usepackage{dsfont}

\newcommand{\method}{SpotLight\space} 

\newcommand{\confPaper}[1]{}

\newcommand{\etal}{\emph{et al.}}

\newcommand{\eg}{\emph{e.g.}}

\newtheorem{Definition}{\hspace{-1em}\bfseries{Definition}}

\providecommand{\mat}[1]{\boldsymbol{\mathrm{#1}}}
\renewcommand{\vec}[1]{\boldsymbol{\mathrm{#1}}}

\DeclareMathOperator*{\argsort}{arg\,sort}

\DeclareMathOperator{\hugeE}{\mbox{\huge\raise-0.3ex\hbox{E}}}
\DeclareMathOperator{\p}{\mathbb{P}}
\DeclareMathOperator{\hugep}{\mbox{\huge\raise-0.3ex\hbox{$\p$}}}

\newcommand{\RR}{\mathbb{R}}

\providecommand{\mX}{\ensuremath{\mat{X}}}

\providecommand{\vs}{\ensuremath{\vec{s}}}

\begin{document}

\title{Insight-centric Visualization Recommendation}

\settopmatter{authorsperrow=4}
\author{Camille Harris}
\affiliation{
\institution{Georgia Tech}
}

\author{Ryan A. Rossi}
\affiliation{
\institution{Adobe Research}
}

\author{Sana Malik}
\affiliation{
\institution{Adobe Research}
}

\author{Jane Hoffswell}
\affiliation{
\institution{Adobe Research}
}
\email{}

\author{Fan Du}
\affiliation{
\institution{Adobe Research}
}

\author{Tak Yeon Lee}
\affiliation{
\institution{Adobe Research}
}
\email{}

\author{Eunyee Koh}
\affiliation{
\institution{Adobe Research}
}

\author{Handong Zhao}
\affiliation{
\institution{Adobe Research}
}

\renewcommand\shortauthors{}

\begin{abstract}
Visualization recommendation systems simplify exploratory data analysis (EDA) and make understanding data more accessible to users of all skill levels by automatically generating visualizations for users to explore. However, most existing visualization recommendation systems focus on ranking all visualizations into a single list or set of groups based on particular attributes or encodings. This global ranking makes it difficult and time-consuming for users to find the most interesting or relevant insights. To address these limitations, we introduce a novel class of visualization recommendation systems that automatically rank and recommend both groups of related insights as well as the most important insights \textit{within} each group. Our proposed approach combines results from many different learning-based methods to discover insights automatically. A key advantage is that this approach generalizes to a wide variety of attribute types such as categorical, numerical, and temporal, as well as complex non-trivial combinations of these different attribute types. To evaluate the effectiveness of our approach, we implemented a new insight-centric visualization recommendation system, SpotLight, which generates and ranks annotated visualizations to explain each insight. We conducted a user study with 12 participants and two datasets which showed that users are able to quickly understand and find relevant insights in unfamiliar data.
\end{abstract}

\ccsdesc[500]{Computing methodologies~Artificial intelligence}
\ccsdesc[500]{Computing methodologies~Machine learning}
\ccsdesc[500]{Mathematics of computing~Approximation algorithms}
\ccsdesc[500]{Information systems~Data mining}

\keywords{
Insight-centric visualization recommendation,
insight-type recommendation,
insight-type ranking, 
insight recommendation
}

\maketitle

\begin{figure}[t!]
\includegraphics[width=1.0\linewidth]{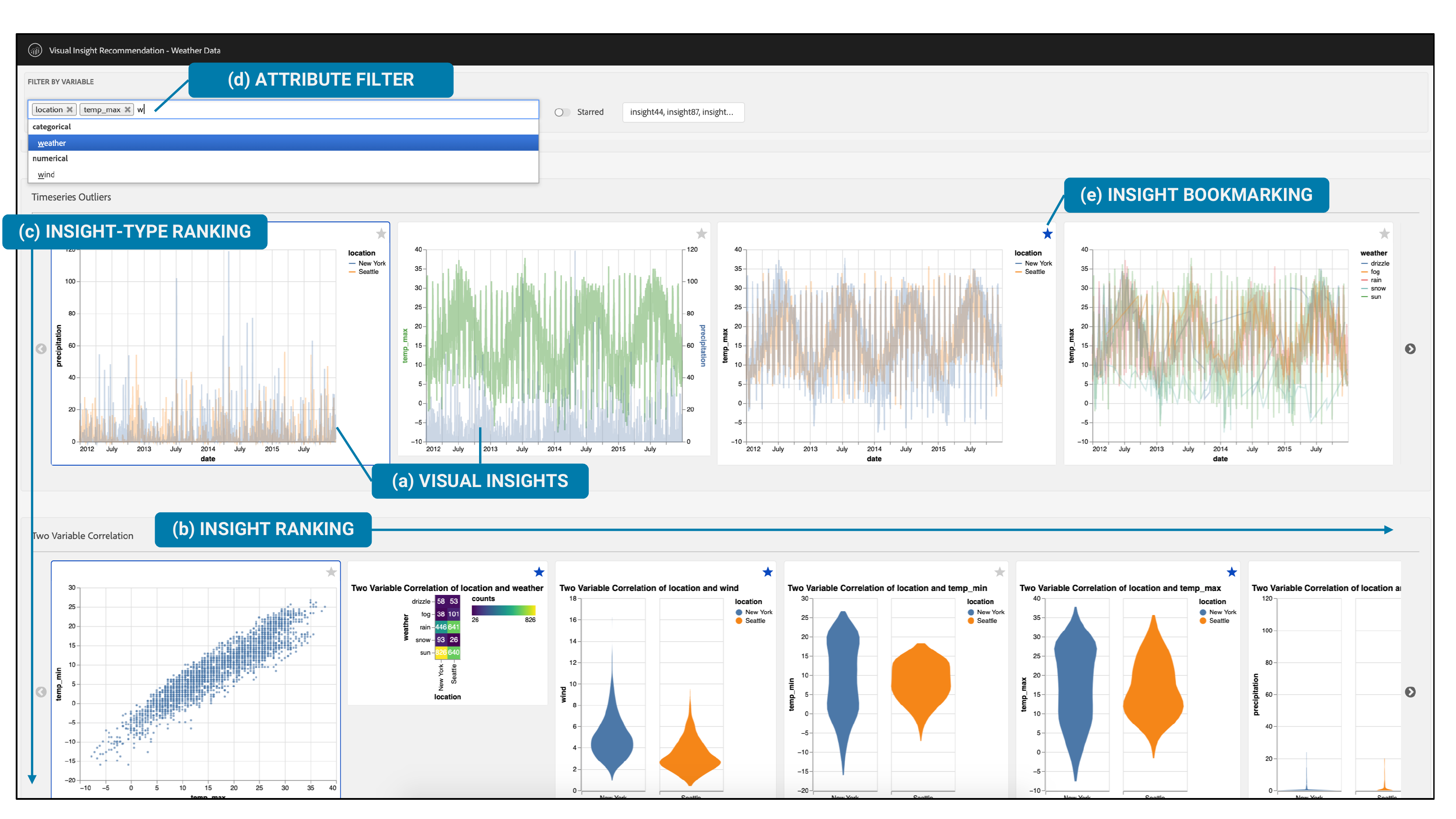}
\vspace{-20px}
\caption{
SpotLight is a \textbf{Visual Insight Recommendation System} that automatically recommends visualizations for the most relevant insights in the data: (a) insights are automatically discovered from a dataset; (b) insights are grouped into rows, scored, and ranked based on their \textit{insight-type}; (c) insight-type rows are globally ranked to show the most impactful insight-types first; (d) users can interactively query the recommendations through attribute filters; and (e) users can bookmark the insights they find most important.\vspace{-15px}
}
\label{fig:overview-data-to-insight-centric-vis-rec}
\end{figure}

\section{Introduction}

Visualization recommendation systems automatically suggest visualization designs to support exploratory data analysis and facilitate the visualization authoring process. These systems generally leverage principles of expressiveness and effectiveness to produce recommendations that are agnostic to the specific data set used. In this paper,
we introduce a novel class of \emph{insight-centric visualization recommendation systems} that are centered around automatically discovered insights from the underlying data.
At the heart of this new class of insight-centric visualization recommendation systems is the notion of automatically discovering, ranking, and visualizing both specific insights and overarching insight-types (or classes of insights) in an arbitrary data set.
Instead of displaying all visualizations together, they are grouped based on meaningful and intuitive insight-types and displayed to the user as rows of visualizations ordered by importance.
The key idea here is that the system scores and recommends the top insight-types, in addition to ranking the individual visualizations within each type.
This ranking enables the user to quickly find the most important and interesting insights.

There are two different but fundamentally important recommendation tasks in the proposed class of insight-centric visualization recommendation systems: 
(1)~recommendation of the important \mbox{top-r} insight-types, and 
(2)~ranking of the \mbox{top-k} insights within each insight-type. In essence, once we obtain the top-k recommended insights, the system can derive the top-r recommended insight-types that best summarize the most valuable insights that were identified in the data. 
Given an arbitrary data set, the proposed class of insight-centric recommendation systems
(i)~automatically discovers the important insights for each insight-type using many different learning and statistical models,
(ii)~scores and recommends the top-r insight-types,
(iii)~scores and recommends the top-k insights for every insight-type,
(iv)~infers an appropriate visualization for each insight,
and (v)~recommends a relevant insight annotation.

Insight-type recommendation enables users to quickly identify the most relevant insight-types in their specific data set, and explore the most important insights within the top ranked insight-types directly. 
Manually finding the appropriate insight-types for a given data set is both costly in terms of the manual effort, but also may require a domain expert with knowledge about the actual data.

To demonstrate the utility of this approach, we contribute a new insight-centric visualization recommendation system: SpotLight.
SpotLight currently supports 21 different insight-types such as skew, multivariate outliers, and nonlinear correlations (Table~\ref{table:insight-discovery-framework}).
The insights for every insight-type are discovered automatically using many different machine learning methods, including both supervised and unsupervised methods. 
Each insight-type is carefully designed to be meaningful to the user, therefore enabling them to quickly understand the relevant insights and their relative importance with respect to any arbitrary user-uploaded data set.

\vspace{-4pt}\subsection*{Summary of Key Contributions} 
\noindent
The main contributions of this work are summarized as follows:

\vspace{5px}\noindent\textbf{Insight-centric Visualization Recommendation.}
We propose a novel class of insight-centric recommendation systems that enables users to quickly understand, explore, and visually identify relevant insights in their data. We further contribute a new system of this type, SpotLight, which automatically discovers, scores, and ranks the insights as well as the insight-types that are most relevant in the data set of interest.
Notably, the most important insight-types and specific insights of each type are displayed to the user in order of importance. We evaluated SpotLight with 12 users to demonstrate the efficacy, and show that users can quickly uncover a diverse and useful set of insights in an unfamiliar data set.

\vspace{5px}\noindent\textbf{Insight-Type Ranking.}
We introduce the notion of insight-type ranking and describe a general approach for recommending the most relevant insight-types for an arbitrary data set.
SpotLight leverages this approach to display the top-r recommended insight-types to the user as rows of visualizations showing insights of that type, with the rows sorted based on the relevance of the underlying insights. The recommendation of insight-types (as opposed to just insights) improves the user experience making it significantly faster and easier for the user to identify the most important insights.

\vspace{5px}\noindent \textbf{Insight Ranking.}
We propose an insight recommendation approach that derives a score 
for each of the discovered insights within an insight-type.
This ranking enables us to derive the top-k insight recommendations for each of the different insight-types. 
A key advantage of our proposed insight ranking approach is that it can naturally leverage multiple insight discovery methods, attribute combinations, attribute types, visualizations, and so on. 

\vspace{5px}\noindent\textbf{Non-Trivial Insights and Insight-Types.} 
The individual insights for every insight-type are discovered automatically using many different machine learning methods, including both unsupervised and supervised techniques.
\method can therefore recommend non-trivial insights that have complex attribute combinations. 
Furthermore, \method naturally handles a wide variety of different attribute types including numerical, categorical, and temporal. 
The recommended insights for a specific insight-type can also have a varying number of attributes and attribute types, thus allowing SpotLight to produce a diverse set of recommended insights.

\section{Related Work} \label{sec:related-work}
In this section, we introduce related work on insight discovery and visualization recommendation, which informs our classification of visualization recommendation systems introduced in Section~\ref{sec:classification}.

\subsection{Insight Discovery}
There has been some work related to insight discovery, focusing both on a specific use case or type of insight, or on the general problem~\cite{10.1145/1168149.1168168}. 
For example, prior work has contributed novel approaches to discovering specific insights including correlations \cite{xiao2017online, wang2017unbiased, wang2005nonlinear, wang2017bagging}, outliers and anomalies \cite{he2003discovering, chandola2009anomaly, gupta2013outlier, duggimpudi2019spatio, duraj2017outlier, liu2017unsupervised}, and peaks \cite{palshikar2009simple}, as well as more general insight discovery techniques \cite{amar2005low, chang2009defining}. Related work often focuses on specific use cases, such as insight discovery for email data~\cite{dey2013email} or geovisualization~\cite{10.1145/2556288.2557228}.
For example, NewsViz \cite{10.1145/2556288.2557228} offers an automatic system to generate interactive geovisualizations for news articles. 
QuickInsights \cite{quickinsights} proposes an algorithm to extract the most useful insights from data and uses
pruning to remove obvious insights and best-first prioritization. 
\mbox{Agrawal et al.~\cite{agrawal1994fast}} propose a general framework for efficiently discovering association rules in data.
Amar et al.~\cite{amar2005low} propose a set of 10 analysis task types which visualization recommendation systems can use to produce effective visualizations. Scorpion~\cite{wu2013scorpion} proposes an approach to reveal qualities in the data that may explain outliers. 
Voder~\cite{voder} helps people explore data through manual view specification. 
Once users create a visualization, Voder utilizes a set of predefined heuristics to generate a list of related data facts (insights). 
DataSite \cite{DBLP:journals/corr/abs-1802-08621}, offers a system that identifies and visualizes data facts based on pre-defined methods.
However, all of this related work focuses on specific insight-types, customized insight discovery methods, or particular data types. In this paper, we contribute a general-purpose solution to insight discovery and ranking, which leverages many of the techniques proposed by related work. Section~\ref{sec:problem-formulation} further outlines the specifics of our problem formalization.

\subsection{Visualization Recommendation}
In this paper, we consider the three primary types of visualization recommendation systems, namely, rule-based~\cite{petajan1997dataspace,wills2010autovis,siddiqui2016effortless,roth1997toward, 4376133,perry2013vizdeck,seo2005rank}, 
\mbox{ML-based~\cite{MLVisRec,vizml, DBLP:journals/corr/abs-1802-08621, kassel2019online, leban2006vizrank, 10.1007/978-3-642-39253-5_60,PVisRec},} 
and insight-based systems~\cite{foresight}. 
Section~\ref{sec:classification} further describes the general system properties considered for this classification.
Related work has also discussed general frameworks for choosing the ideal visualizations for specific data \cite{few2017data} and providing a general overview of how to approach visualization recommendation based on other recommendation system techniques~\cite{vartak2017}.

\textsc{\it Rule-based Systems.}
Most previous work on visualization recommendation has focused on rule-based systems~\cite{viegas2018generating,10.1145/108360.108361,vartak2015seedb} where visualizations are recommended based on a series of manually defined if-then statements~\cite{4376133,voyager,perry2013vizdeck,seo2005rank}. 
Voyager \cite{voyager} proposes a hybrid approach which incorporates rule-based decisions with support for user selected variables (columns). 
In addition, some related work has proposed different types of visualization generation tools \cite{petajan1997dataspace,siddiqui2016effortless,wills2010autovis} as well as tools to help users produce their own visualizations, such as VizAsisst~\cite{bouali2016vizassist} and Tableau \cite{information_tableauversion._2016}.

\textsc{\it ML-based Systems.} 
VizML~\cite{vizml} focus entirely on design choice classification (\eg, predicting the chart type),
and not other more challenging problems of visualization recommendation, such as selecting the appropriate variables for the visualization.
Data2Viz~\cite{data2vis} uses an end-to-end neural translation model to map data specifications to visualizations. 
DeepEye \cite{deepeye} combines rule-based visualization generation techniques similar to work in the previous paragraph with classification models to rank visualizations and classify them as "good" or "bad." 
DataShot \cite{datashot} creates a data fact sheet by analyzing the data for the most interesting data facts and generating visualizations from them. 
Draco~\cite{draco} extends the rule-based system of Voyager by learning weights for manually defined constraints.
More recently, Xin et al.~\cite{MLVisRec} proposed an end-to-end learning-based visualization recommendation model that is trained using a large corpus of hundreds of datasets and visualizations.
Afterwards, the learned model can be applied to recommend visualizations given any arbitrary new unseen dataset of interest.
Other recent work has focused on learning a personalized visualization recommendation model called PVisRec~\cite{PVisRec} that can recommend interesting and highly relevant visualizations for a given user based on their past data and visualization preferences.
However, none of this related work focuses on recommending insights, and thus solves a different problem.

\textsc{\it Insight-based Systems.}
Insight-based systems recommend visualizations based on properties of the specific data set that will be visualized. Our system SpotLight is an example of this type. This space has largely been under explored, but the work closest to our own is Foresight~\cite{foresight}. Foresight is an early example of an insight-based system that employs a single heuristic insight method to identify insights based on the marginal distribution of the attributes. Foresight then selects a manually defined visualization based on the insight-type. In contrast, we propose a more general approach to insight-centric recommendation and ranking, and instantiate the techniques in our own system, SpotLight. SpotLight automatically creates a novel two-dimensional ranking of insight-types and individual insights, that leverages many different learning-based insight discovery methods to produce a robust ranking. Furthermore, our proposed approach generalizes to a wider variety of attributes types and a diverse set of visualization designs.
Notably, this work contributes a novel approach for ranking and recommending both individual insights and overarching classes of insight-types to facilitate rapid identification of relevant insights in an arbitrary \mbox{data set.}

\section{Classes of Recommendation Systems} \label{sec:classification}

In this paper, we introduce a novel class of \emph{insight-centric} visualization recommendation systems, that goes beyond the existing \emph{design-centric} and \emph{preference-centric} classes explored in prior work. As a whole, visualization recommendation systems aim to automatically generate visualizations, often based on some form of user input. However, each specific class of visualization recommendation system employs unique characteristics to control the recommendation process. In this section, we describe the key differences between the classes of visualization recommendation systems and propose the unique benefits posed by insight-centric approaches.

\vspace{6px}
\noindent\textbf{Design-centric systems} leverage information about the underlying data types (\eg, nominal, ordinal, quantitative) to recommend visualizations based on principles of visualization effectiveness and expressiveness. The goal in such systems is to create ``good'' visualization designs by default to help users rapidly create visualizations. However, such systems are based entirely on principles of visualization design, and remain agnostic to the underlying data distributions aside from a basic consideration of data types. 

\vspace{6px}
\noindent\textbf{Preference-centric systems} go one step further to incorporate new knowledge that has been learned about the visualizations and data preferences. Such systems may employ machine learning techniques to assign weights based on properties like the underlying chart and/or data types, or to create a global ranking of visualization designs. While details of the underlying data set may surface during this process, the data itself is still not considered directly.

\vspace{6px}\noindent\textbf{Insight-centric systems} aim to incorporate details of the specific underlying data that will be used to create the visualizations. Whereas the other classes of visualization recommendation systems are largely agnostic to the underlying data values, insight-centric systems incorporate the specific data properties as an essential feature of the recommendation process. Such systems can therefore highlight not only the most relevant visualizations, but also summarize the most important classes of insight-types to provide a high-level overview of the shape of the underlying data. We provide additional details about the exact problem formulation for insight-centric recommendation systems in the subsequent section.

\definecolor{googlepurple}{RGB}{128,0,128}
\definecolor{googlepurple}{RGB}{121,28,180}

\section{Problem Formulation} \label{sec:problem-formulation}
An insight is a new piece of knowledge identified in the data while performing an analytical task. In other words, an insight is a property of the data that is unexpected, complex, deep, or relevant to the analyst, such as a strong linear correlation between two variables or a set of points that are anomalies with respect to time. Providing insights is arguably the main goal of information visualization~\cite{chang2009defining}. However, none of the existing visualization recommendation systems described in prior work focus on the recommendation of classes of insight-types or the individual ranking of visual insights. 

\begin{Definition}[Visual Insight Recommendation]\label{def:visual-insight-rec}
\; \; \; \; \;
Let $\mathcal{I} = \{I_1, I_2, \ldots, I_{|\mathcal{I}|}\}$ denote the set of insight classes. 
Given an arbitrary user data set $\mX$,
we define $\mathcal{F}_{I_i} = \{f_1, f_2, \ldots\}$ as the set of insight discovery methods for insight class $I_i \in \mathcal{I}$ and $\rho_{I_i}(\mX_1,\mX_2,\ldots)$ as the set of insights found for that class.
An insight-centric visualization recommender system 
(i) automatically discovers the important insights for each insight class $I_i \in \mathcal{I}$ using many different learning and statistical models $\mathcal{F}_{I_i}$ for many different attribute types and attribute type combinations;
(ii) recommends the important top-r insight classses; 
(iii) scores and recommends the top-k insights $\rho_{I}(\mX_1,\ldots,\mX_k)$ for every insight class $I_i \in \mathcal{I}$ and attribute type combination;
and (iv)~infers an appropriate visualization for each insight.
\end{Definition}
\noindent
Every insight class $I_i \in \mathcal{I}$ has a set of attribute type combinations denoted as $\mathcal{A}_I$. For each attribute type combination $A \in \mathcal{A}_I$, we have a set of insight discovery methods $\mathcal{F}_{I}$ as shown in Table~\ref{table:insight-discovery-framework}.
As an example, the "Two Variable Outliers" class has two different attribute type combinations $\mathcal{A}_I = \{N \times N, C \times C\}$. For each insight class and each attribute combination, we use many different methods to detect the important insights; for example, DBScan and IForest are used among others for detecting "Two Variable Outliers" insights with attribute type combination $N \times N$.
Furthermore, we also recommend many different visualizations for every insight type $I \in \mathcal{I}$, as shown in Table~\ref{table:insight-discovery-framework}.

\begin{table}[t]
\centering
\caption{
Auto-Insight Discovery Framework.
We summarize the insight-types, and corresponding attribute type combinations and chart types supported for each insight. We also summarize the techniques used in the proposed framework for automatically revealing these insights, and color each one based on the approach used for discovering the insights 
(\textcolor{googlered}{\bf information theoretic}, \textcolor{googlegreen}{\bf statistical}, \textcolor{googleblue}{\bf supervised \emph{\bf learning}}, \textcolor{googlepurple}{\bf unsupervised \emph{\bf learning}}). 
$N$=numerical, $C$=categorical, $T$=time.}
\label{table:insight-discovery-framework}
\vspace{-2mm}
\includegraphics[width=\columnwidth]{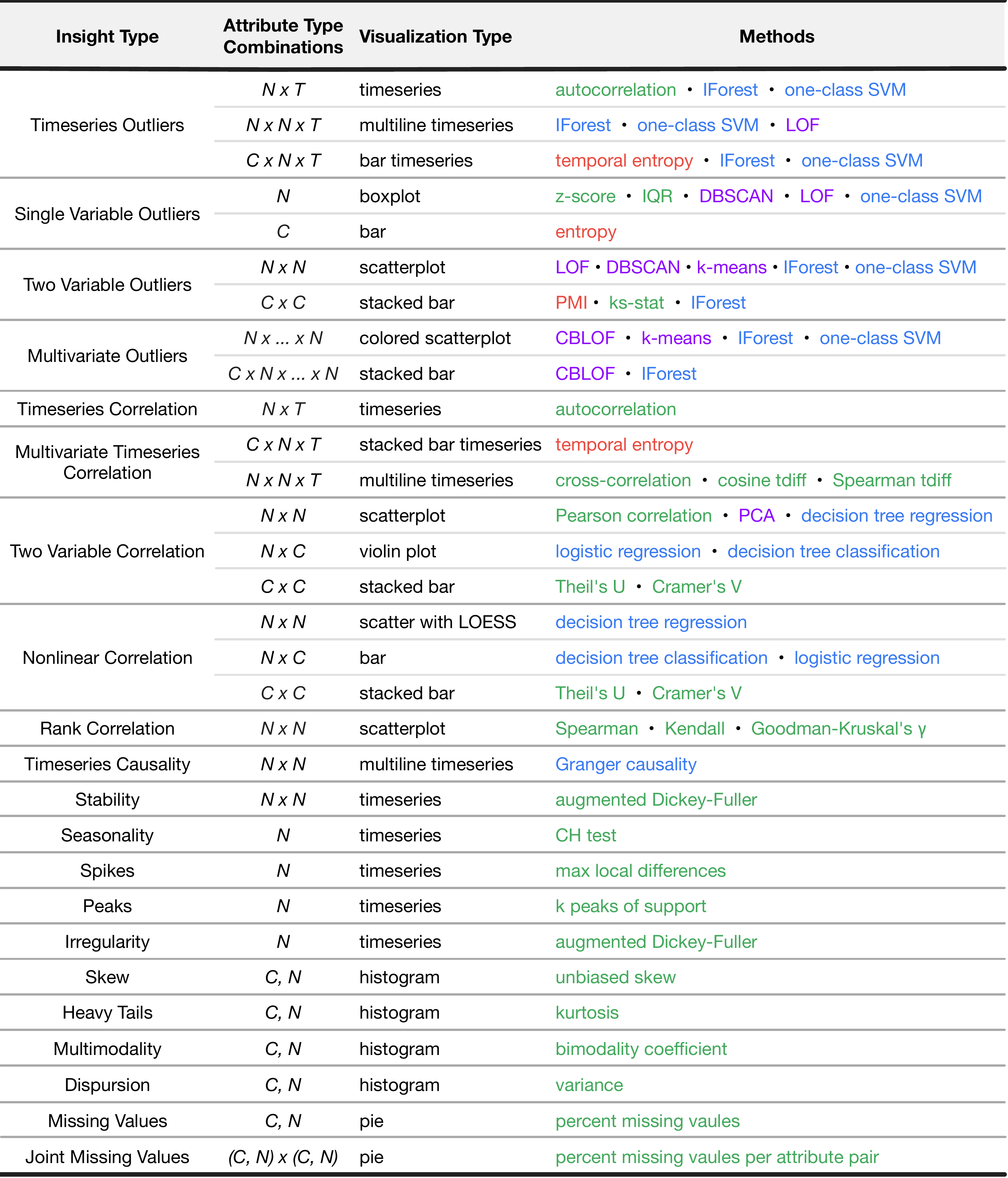}
\vspace{-25px}
\end{table}

\section{Visual Insight Recommendation} \label{sec:approach}
Given this problem formulation, we now describe our proposed approach for visual insight recommendation.

\subsection{Auto-Insight Discovery} \label{sec:auto-insight-discovery}

The first step in the recommendation process is to automatically detect insights in an arbitrary dataset. To effectively identify insights for any input dataset, we employ a variety of different insight discovery methods as described in the following sections.

\subsubsection{Multiple Insight Discovery Methods} \label{sec:multiple-insight-discovery-methods}
Given an arbitrary dataset, each insight type (and attribute combination) has many potential algorithms and the best algorithm depends on the data and its characteristics~\cite{wolpert1997no,wolpert1995no}. 
The above is due to the no-free lunch theorems by Wolpert~\etal~\cite{wolpert1997no,wolpert1995no}.
Intuitively, the no-free lunch theorem states that any two optimization algorithms are equivalent when averaged across all possible problems (datasets)~\cite{ho2002simple,joyce2018review}.
Since the ``best'' insight discovery method depends on the data and its characteristics~\cite{domingos2012few}, we instead use multiple insight discovery methods for each insight-type and attribute combination.
This ensures that our approach is able to find important insights, regardless of the user selected dataset and its underlying characteristics.
Hence, our approach can perform effectively for any user uploaded dataset, as opposed to a very small set of possible datasets.

\subsubsection{Learning-based Insight Discovery Methods} \label{sec:learning-based-insight-methods}
In addition to the simple heuristic-based insight methods used in Foresight (\eg, IQR), we use learning-based insight methods for many of the different insight-types as shown in Table~\ref{table:insight-discovery-framework}.
In Table~\ref{table:insight-discovery-framework}, we summarize the different insight types, the attribute combinations supported for each insight-type, the visualizations and chart-types used for every insight-type and attribute combination, along with the machine learning insight discovery methods used for the different insight-types and attribute combinations.
Many of the machine learning methods in the framework can be configured with different kernel and distance functions such as One-Class SVM, Kernel PCA, k-means, among many others.
These methods also have many other real-valued hyperparameters that can be optimized.
Therefore, we sample a few models from each to use in the ranking of the different insights (\eg, One-Class $\{\text{linear}, \text{RBF}, \text{polynomial}\}$-SVM).

\subsubsection{Insight-Type Granularity}
As the number of different insight-types in visual insight recommendation systems becomes large, the insight-type ranking and recommendation that this work proposes becomes even more important.
Intuitively, the more insight-types the visual insight recommendation system incorporates, the higher the probability of finding relevant insights that are important to the user and dataset of interest.
SpotLight also detects insights at a fine-granularity that are more meaningful and intuitive to the user.
For instance, instead of considering a general insight-type called ``outliers'' as done in~\cite{foresight}, we distinguish outliers into meaningful subcategories, such as time-series outliers, multivariate outliers, single variable outliers, and so on (Table~\ref{table:insight-discovery-framework}).

\subsubsection{Multi-Attribute Types and Combinations} \label{sec:multi-attr-types-and-combinations}
For each insight-type shown in Table~\ref{table:insight-discovery-framework}, we consider multiple attribute combinations and many different attribute types per insight-type.
As an aside, other work such as~\cite{foresight} only considers a single attribute combination and only support numerical attributes (as opposed to numerical, categorical, and temporal considered by our proposed system).

\subsubsection{Multi-faceted Insight Types}
Many insight-types support different attribute combinations ($N \times C$, $C \times C$), as well as combinations consisting of a different number of attributes, \eg, $N \times N$ and $N \times N \times C$. 
One such example in Table~\ref{table:insight-discovery-framework} is timeseries outliers.

\begin{figure*}[t!]
\vspace{-4mm}
\centering
\includegraphics[width=0.9\linewidth]{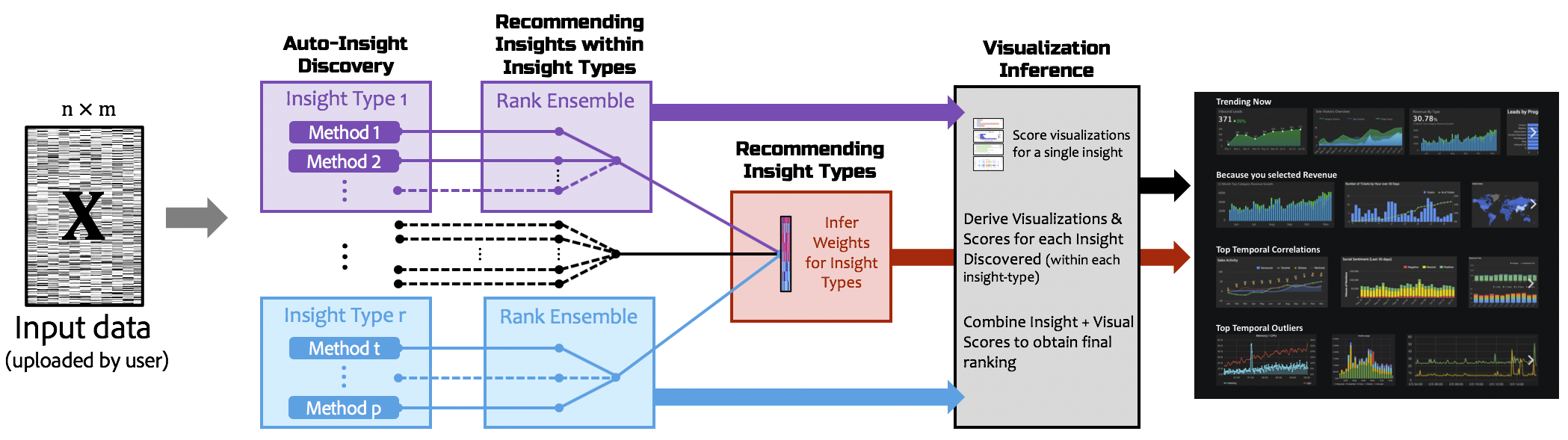}
\vspace{-2mm}
\caption{Overview of the proposed class of Insight-Centric Visualization Recommendation Systems.}
\label{fig:approach-overview}
\vspace{-4mm}
\end{figure*}

\subsubsection{Methods} \label{sec:methods}
In this work, we use many different methods to automatically discover insights for each insight-type.
We provide a summary of these methods in Table~\ref{table:insight-discovery-framework}.
We categorize these methods into four main classes, including 
(i)~\textcolor{googlered}{\bf information theoretic}, 
(ii)~\textcolor{googlegreen}{\bf statistical}, 
(iii)~\textcolor{googlepurple}{\bf unsupervised learning}, and
(iv)~\textcolor{googleblue}{\bf supervised learning}.
We adapt and leverage many of these methods in nonstandard ways for learning and automatically discovering insights from a wide range of different insight-types.
To do this, we often have to first derive a different representation of the data, modify or change entirely the scoring function, etc.
As such, we focus our discussion on the fundamentally different and novel ways these techniques were used in our insight recommendation framework.
Importantly, the proposed insight recommendation framework is flexible as the insight discovery methods used for each insight-type and attribute type combination are completely interchangeable.
Furthermore, this approach can generalize to any arbitrary set of methods that users may want to employ in the future.

\subsection{Insight Ranking} \label{sec:ranking-insights-within-insight-type}
Once insights have been discovered, the next step is to create a ranking of insights to surface for the user.
Previous work used a single heuristic for each insight-type and therefore did not require any scoring or ranking metric.
In contrast, our approach automatically reveals insights of a specific type using multiple learning-based methods, which makes scoring and ranking the insights from across multiple methods non-trivial.
Ranking becomes even more complex given that particular insights from an insight-type can leverage a different number of attributes of different types and combinations as shown in Table~\ref{table:insight-discovery-framework}, which makes comparing and scoring individual insights within a particular insight-type difficult.

Given an insight-type $I \in \mathcal{I}$ and set of methods $\mathcal{F}_{I}$ for that insight type, the goal is to assign a score to each insight detected using $\mathcal{F}_{I}$.
Suppose $p_I = |\mathcal{F}_{I}|$, then we may have $p_I$ scores for every insight detected for insight type $I \in \mathcal{I}$.
We define the utility function $\phi$ as:
\begin{equation}
\phi : \mathcal{X} \times \mathcal{F}_I \rightarrow \RR
\end{equation}\noindent
where $\mathcal{X}$ is the space of attribute combinations used by the methods $\mathcal{F}_I$ for insight-type $I \in \mathcal{I}$.
We can derive a final score as follows:
\vspace{-4px}
\begin{equation}\label{eq:general-ranking-formulation}
\phi(\mX_{k}, \mathcal{F}_{I}) = \frac{1}{Z} \sum_{f_i \in \mathcal{F}_{I}}\; \sum_{j=1}^{n} \;\big[g\big(f_i(\mX_{k},\Lambda_i)\big)\big]_{j}
\end{equation}
\noindent where $\mX_{k}$ is an arbitrary attribute matrix consisting of one or more attributes from $\mX \in \RR^{n \times m}$.
In other words, $\mX_{k}$ is an attribute combination matrix that may consist of one or more attributes.
Further, $\big[g\big(f_i(\mX_{k})\big)\big]_{j}$ is the $j$th value from $\vs = g\big(f_i(\mX_{k})\big) \in \RR^{n}$.
We define $g : \RR \rightarrow [0,1]$ and $\Lambda_{i}$ as the set of hyperparameters for the learning-based insight method $f_i \in \mathcal{F}_{I}$.
Suppose $f_i \in \mathcal{F}_{I}$ is one-class SVM, then $\Lambda_i$ may include a kernel function $\mathbb{K}$ such as the non-linear RBF kernel or polynomial kernel along with other hyperparameters, such as $\gamma$ or the degree of the polynomial kernel. 
In Eq.~\ref{eq:general-ranking-formulation}, $Z = |\mathcal{F}_{I}|n$.

Our formulation in Eq.~\ref{eq:general-ranking-formulation} assumed that the output of each insight method was the same. 
However, in general, some insight methods may return scores for only the most relevant data points (as opposed to all $n$ data points), or even a single score for the attribute combination.
In this case, we can rewrite Eq.~\ref{eq:general-ranking-formulation} as follows:
\vspace{-4px}
\begin{equation}\label{eq:general-ranking-formulation-nonuniform-output}
\phi(\mX_{k}, \mathcal{F}_{I}) = \frac{1}{|\mathcal{F}_{I}|} \sum_{f_i \in \mathcal{F}_{I}}\; \frac{1}{n_i} \sum_{j=1}^{n_i} \;\big[g\big(f_i(\mX_{k},\Lambda_i)\big)\big]_{j}
\end{equation}
\noindent where $n_i$ denotes the number of scores returned by $f_i$.

Eq.~\ref{eq:general-ranking-formulation-nonuniform-output} assigns an insight score to the attribute combination $\mX_k$ for insight type $I \in \mathcal{C}$ using the methods $\mathcal{F}_{I}$.
Now, we can obtain a ranking of the insights within the insight type $I \in \mathcal{C}$ as follows:
\vspace{-4px}
\begin{equation} \label{eq:ranking-insights-within-insight-type}
\rho_{_{\!I}}\big(\{\mX_{1},\ldots,\mX_k,\ldots\}\big)\, =\; \argsort_{k} \; \phi(\mX_{k}, \mathcal{F}_{I})
\end{equation}
\vspace{-6px}

There are a number of important and key novel aspects in the above formulation for ranking insights within an insight-type.
More specifically, we are the first to 
(i) leverage learning-based insight methods,
(ii) use more than a single method per insight,
(iii) propose and require an actual insight scoring and ranking function,
among many other fundamental differences. Moreover, we describe a general mathematical framework for recommending visual insights. 

Instead of using the normalized scores from every method to derive a final score for a given insight-type and attribute combination, we derive a final score for each data point (row of $\mX_{k}$) based on the rankings given by each method.
Using the ranks as opposed to the weights can help avoid biasing certain methods depending on the distribution of inferred weights.
For a single $j$ and attribute combination $\mX_k$, we can generate a rank-based score as follows:
\vspace{-4px}
\begin{equation}\label{eq:general-rank-based-ranking-formulation}
R_j(\mX_{k}, \mathcal{F}_{I}) = \frac{1}{|\mathcal{F}_{I}|} \sum_{f_i \in \mathcal{F}_{I}}\; \pi_j \big(f_i(\mX_{k},\Lambda_i) \big)
\end{equation}\noindent
where $\pi_j \big(f_i(\mX_{k},\Lambda_i) \big)$ is the position of the $j$th object in the ranking obtained from attribute combination $\mX_k$ with method $f_i \in \mathcal{F}_{I}$.
Therefore, $R_j(\mX_{k}, \mathcal{F}_{I})$ is the average rank of $j$ across all methods
and every method can be seen as having equal weight.
Notice that Eq.~\ref{eq:general-rank-based-ranking-formulation} gives an overall ranking for each data point $j$ (row in $\mX$) whereas Eq.~\ref{eq:ranking-insights-within-insight-type} provides a ranking of the overall attribute combinations $\{\mX_1,\ldots,\mX_k,\ldots\}$ across all $n$ data points and $|\mathcal{F}_{I}|$ methods. 

The rank-based score given by Eq.~\ref{eq:general-rank-based-ranking-formulation} can be used to appropriately annotate the visualizations for better visual insight recommendation.
As an example, suppose we use the set of two variable outlier methods and use the methods as an ensemble to obtain an overall ranking of the data points by how much of an outlier each point appears.
This can be accomplished using the average rank of the data points given by the set of methods.

Note that if $g$ is set to the min-max norm for each insight-type and attribute type combination, then by definition we are guaranteed to have a diverse ranking of visualizations for each insight-type. 
Intuitively, since min-max norm is applied to each attribute type combination independently, then one of the insights with that attribute type combination is guaranteed to score 1.
Hence, if there are three attribute type combinations for a given insight-type, then the first three insights will be of different attribute type combinations.

\subsubsection{Space \& Scalability}
To ensure the approach is space-efficient and scalable for large datasets with millions or more data points, we do not store the scores of each data point (for all methods).
Notably, we can simply compute the final score on-the-fly while taking a scan over the (sampled) attribute values.
For large data where even a linear scan over the data points is considered too costly, we can use sampling or sketching techniques.

\subsection{Scoring and Ranking Insight Types} \label{sec:insight-type-ranking}
The insight-types are scored in a completely automatic and data-driven fashion.
In particular, the ranking of the different insight-types (rows of insights and their visualizations) is driven by the amount of information captured by each insight-type across all the methods used for discovering insights of that insight-type.
Let $Q_I = \{\mX_{1},\ldots,\mX_k,\ldots\}$ denote the set of potential insights being scored for insight type $I \in \mathcal{I}$, then
\begin{equation} \label{eq:scoring-insight-types}
\Psi(I) \;=\; \frac{1}{|Q_I|} \sum_{\mX_k \in Q_I} \phi(\mX_{k}, \mathcal{F}_{I})
\end{equation}\noindent
where $\Psi(I)$ is the overall score assigned to insight type $I \in \mathcal{I}$ for the dataset as a whole.
As defined by  Eq.~\ref{eq:scoring-insight-types}, if an insight-type $I \in \mathcal{I}$ receives a relatively high score $\Psi(I)$, then there must be many important and highly weighted insights of that insight type. 
The overall insight-type score is derived in a completely automatic and data-driven fashion based on the insights discovered in the specific dataset and the scores of those insights derived previously in Section~\ref{sec:ranking-insights-within-insight-type}.
Using Eq.~\ref{eq:scoring-insight-types}, we derive a global ranking of the insight types $\mathcal{I}$ 
by simply sorting them based on their overall scores:
\begin{equation} \label{eq:ranking-insight-types}
\rho_{_{\!\mathcal{I}}}\big(\{I_{1},I_{2},\ldots,I_{|\mathcal{I}|}\}\big)\, =\; \argsort_{I \in \mathcal{I}} \; \Psi(I)
\end{equation}\noindent
where for any two insight types $I_{i}$ and $I_{j}$ in the insight-type ranking
$\rho_{_{\!\mathcal{I}}}\big(\{I_{1},I_{2},\ldots,I_{|\mathcal{I}|}\}\big)$ 
such that $i<j$, then $\Psi(I_i) \geq \Psi(I_j)$ holds by definition.
Hence, the insights of insight-type $I_i$ are displayed before $I_j$, since the insight-type score $\Psi(I_i)$ of $I_i$ is larger indicating that there are more important and useful insights of insight-type $I_i$ in this specific dataset relative to the other insight-type $I_j$.
The insight-types are then displayed to the user according to the ranking from Eq.~\ref{eq:ranking-insight-types}.
This ranking enables the user to quickly find the most relevant insights for the dataset of interest.
Furthermore, the overall insight-type ranking for a given dataset can be used to better understand the data quickly, \eg, if the top most important insight-types for a specific dataset are all related to time-series (such as time-series outliers, time-series causality, and so on), then the user immediately knows that the temporal dimension in the data is important.

\subsection{Discussion}
We now discuss a few other important components of SpotLight.

\subsubsection{Multiple Visualizations per Insight-Type} \label{sec:vis-inference}
The next step is to derive weights (either learned or rule-based) for the potential visualizations for each of the top-k insights of a specific insight-type learned by our proposed approach.
For each insight, we derive weights 
for the potential visualizations and select the best visualization (rank-1).
With this approach, notice that we could easily select and display more than one visualization for a given insight.
This option may be preferable if there is an insight that is significantly more important than others.
In this case, showing several different visualizations for the same insight may be more useful to the user than showing less relevant insights or visualizations.

\subsubsection{Data Complexity}
Given an insight from an arbitrary insight-type, we add a penalty to overly complex and data-intensive insights.
For instance, an insight involving time, a categorical attribute, and two numerical attributes will obviously be more difficult for a user to understand and visualize than one involving only two attributes.

\subsubsection{Human-in-the-Loop}
The system also leverages intuitive UI features that allow the user to specify attributes of interest and perform many other types of interactive insight queries as well.
Importantly, the recommended insight-types and visual insights within each type are automatically adapted based on the user input such as the attributes of interest.

\section{Evaluation}\label{sec:exp}
To understand the efficacy, accuracy, and generalizability of the proposed class of insight-centric recommendation systems, we conducted a user study of the system with twelve participants. 
The user study is designed to answer the following questions:
\begin{compactitem}
\item[\bf RQ1:] Does SpotLight help users generate hypotheses, find insights, and  understand the data better?
\item[\bf RQ2:] Are the supported insight-types and attribute types useful, diverse, and comprehensive?
\item[\bf RQ3:] Are the recommended rankings (of both insight-types and insights) useful and do they provide diversity?
\item[\bf RQ4:] Are the visualizations used for the recommended insights useful, diverse, aesthetic, and easy-to-understand?
\item[\bf RQ5:] Does the UI allow users to find the most interesting insights and does it support different workflows for EDA? 
\end{compactitem}\noindent

\subsection{Method} \label{sec:exp-user-study-procedure}

\noindent\emph{Participants.} \label{sec:exp-user-study-participants} Participants were recruited by word-of-mouth and internal mailing lists at a software company and two universities. In total, there were 12 participants aged 25--51 years ($mean=34$). Participants consisted of two graduate students, five researchers/data scientists, four technical professionals, and one visualization designer. They had 3--20 years of data analysis experience and 1--20 years visual analytics experience ($mean=8$ and $7$ respectively). Two participants were female and ten were male.\\

\noindent\emph{Procedure.} Sessions were conducted remotely, through video conference and screenshare, with two researchers (one administered, one took notes). Participants were first trained on SpotLight's UI and features using a training dataset. Participants were then asked to perform a free-form exploration of two datasets
from the UCI ML Repository: 
\texttt{Weather} data 
and
\texttt{Wholesale} customer profile data. 
No participant was familiar with either dataset ahead of time and the order of the datasets was counterbalanced between participants to avoid fatigue and learning effects.
During exploration, participants were encouraged to think-aloud and bookmark interesting charts that helped them understand the data. 
After exploring the data, participants were asked to verbally present their findings.  At the end of the session, participants completed a post-questionnaire with twenty-nine 7-point Likert questions and three questions about the usability and utility of SpotLight. The bookmarked insights were also logged for each participant. Participants were required to use a desktop or laptop computer instead of mobile or tablet devices, for a consistent experience. Each session lasted 60--75 minutes.\\

\noindent\emph{Datasets.} The \texttt{Weather} dataset consisted of weather patterns for two US cities, Seattle and New York, over four years (2014--2017). Each row represented one day for each city, with measurements for: precipitation (\textit{in}), wind (\textit{mph}), minimum and maximum recorded temperatures (\textit{C}). Lastly, there was a categorical variable indicating the type of weather seen on that day, such as \textit{sunny, fog, drizzle, rain,} or \textit{snow}. The \texttt{Wholesale} dataset consisted of 440 businesses' annual spending at a wholesale company. Each row represented one customer, with a categorical variable representing the type of business (\textit{hospitality} or \textit{retail}) and six numerical values representing annual spending broken down by department: \textit{fresh, frozen, grocery, detergent and paper, dairy}, and \textit{deli}.\\

\noindent\emph{Analysis.} \label{sec:exp-user-study-analysis}
Sessions were analyzed by two researchers to observe participants' patterns in workflow and their comments about their use and understanding of the system. 
The insights users bookmarked were saved and analyzed based on frequency and ranking in the recommendations. 
In particular, since the insights within every insight-type are ordered automatically by SpotLight, we compared the ranking of insights given by SpotLight to the ``ground-truth'' ranking given by the insights users bookmarked.
Furthermore, we used the bookmarked insights to quantitatively evaluate the insight-type ranking derived by SpotLight compared to the ranking of insight-types given by the bookmarked insights.
Finally, the post-questionnaire Likert responses were analyzed with descriptive statistics and the open-ended responses were thematically coded.

\subsection{System Usage} 
To gather preconceptions about the data and to confirm their understanding of the datasets, we asked participants what they expected or would find interesting to see in each of the datasets. Participants tended to have more hypotheses about the weather dataset than the wholesale dataset because it was more natural to have experience with weather. For example, nine participants hypothesized city-based comparisons (e.g., “Seattle is rainier than New York”) and half of the participants were interested in variable-based correlations (e.g., whether temperature and precipitation have a relationship). Additionally, the types of hypotheses people had were different based on the dataset. Wholesale hypotheses were less focused on comparisons and were more related to the relationships between different spending variables in general, though some did also focus on comparing customer types (i.e., hospitality vs. retail). 

The hypotheses also impacted participants’ workflow when using the system. In general, most participants (10/12) followed a general exploratory workflow: they began by reviewing the variable in the dataset using the variable search bar then explored the recommended insights row-by-row. Most opted to explore the results from top to bottom, citing trusting the recommendations. \textit{``Since the system suggested this as the most insightful, I’ll go ahead and start with that''} (P1). After reviewing the top recommendations and after generating additional hypotheses, some participants filtered the results based on specific variables there were interested in analyzing. Conversely, many participants (5/12) with clear hypotheses began their analysis by first filtering by the variables relevant to their ideas to find insights that confirmed or denied their hypotheses. Two participants (P1, P5) systematically searched for variables one-by-one to analyze subgroups of the results and to understand the dataset better. \textit{"Maybe I need to first look at one variable at a time, so I can understand the dataset first. It’s a lot cleaner and easier to see initially with one variable"} (P1). 

\subsection{Insight and Insight-type Results} 
We analyzed the rank of users' bookmarked insights to quantitatively evaluate the effectiveness of the insight and insight-type rankings of our proposed approach.

\smallskip\noindent\emph{Insight Ranking.}
Across all insight-types, 68\% (weather) and 61\% (customer wholesale) of the insights bookmarked by the users came from the top-5 insights recommended by SpotLight in their respective insight-type rows.
This finding indicates that the ranking of insights (within each insight-type) from SpotLight were indeed useful and validates the effectiveness of the ranking of insights and the top-recommended insights within each insight-type given by the proposed approach.
Considering the top-10 insights recommended by SpotLight (as opposed to only the top-5), we find that among all the bookmarked insights from the participants, 91\% (weather) and 86\% (customer wholesale) of them can be found in the top-10 ranked insights from SpotLight.
This is important, since there is clearly an exponential amount of visual insights that can be recommended for every different insight-type, and the above finding demonstrates that SpotLight is able to accurately score and recommend the best insights with the highest quality/relevance from each insight-type.
In addition, users bookmarked multiple chart-types per insight-type, indicating that the diversity of charts was useful.
This is particularly important since the proposed insight ranking function biased charts of different types towards the top of the ranking.
We posit that this was welcomed by the users for many reasons.
For instance, if they don't understand certain chart types, then they can immediately see the variety of different charts used for the insight-type.
Furthermore, if there was a certain type of chart that the user found especially useful for a given insight-type, then they could quickly find the insights that had such a chart.

\smallskip
\noindent\emph{Insight-Types.}
We leverage the insights bookmarked by the users to 
(i) investigate the utility of the insight-types, and 
(ii) quantitatively evaluate the \emph{insight-type ranking}.
We begin by first evaluating whether the automatic insight-type ranking proposed in this work is important or not.
To do this, we compare the top-5 insight-types ranked by the number of insights bookmarked from each insight-type across both weather and customer wholesale data.
We use Kendall's $\tau$ rank correlation to compare the top-5 insight-types from either dataset.
Strikingly, we find the ranking of the top-5 insight-types for weather data are significantly different at \texttt{p-val}=0.05 from those preferred by participants in the customer wholesale data.
Hence, ranking the insight-types is indeed useful, since the insight-types for either dataset were ranked differently by the users.
In other words, if the ranking of insight-types did not matter, then the ranking should not depend on the dataset of interest, and thus be the same.
This result indicates the importance of ranking the insight-types differently depending on the user-selected dataset, the important insights discovered in the data, as well as the behavior/preferences of the user.
In addition, we also found that users bookmarked insights from every insight-type but one (heavy tails).
This result holds across both datasets and indicates that nearly all insight-types proposed in SpotLight are useful.

\smallskip
\noindent\emph{Insight-Type Ranking.}
To quantitatively evaluate the effectiveness of the insight-type ranking, we compare the ranking of insight-types from SpotLight to the ranking given by the participants (based on their bookmarked insights) for the wholesale data.
The ranking of insight-types from the participants are used as ground-truth.
To test whether the insight-type ranking given by SpotLight is similar to the ranking given by the participants in the study, we use Kendall's $\tau$ rank correlation.
Importantly, we find the insight-type ranking from SpotLight to be significantly correlated with the ground-truth ranking from the users in the study (with a pval<0.01).
Notably, the rank correlation coefficient $\tau=0.87$ indicating the two rankings are extremely correlated with a p-value of 0.0008.
This result quantitatively demonstrates the effectiveness of the insight-type ranking as the ranking given by SpotLight is correlated significantly with the user-preferred insight-type ranking.

\begin{table}[!t]
\vspace{-1mm}
\centering
\caption{
Overall ratings of SpotLight on 7-point Likert scale. 
}
\label{table:user-study-results}
\vspace{-4mm}
\centering 
\small
\setlength{\tabcolsep}{5.3pt} 
\def\arraystretch{1.0} 
\begin{tabular}{
@{}
rc
H
cc
H
@{}
}
\midrule
\textsc{Assessment} & \textsc{Measurement} & \textsc{Question} & \textsc{Mean} & \textsc{SD} &   \\
\midrule
\multirow{2}{*}{\textbf{System} (RQ1)} 
& Usability & \emph{I understood the datasets better after using this system.} & 6.25 & 0.87 &  \\
& Useful & \emph{I found the system to be useful.} &  5.92 & 0.51 &  \\
\midrule
\multirow{3}{*}{\textbf{Insight-types} (RQ2)} 
& Useful & \emph{The insight types used in the system were useful} & 5.58 & 1.16 &  \\
& Diversity & \emph{The insight types used in the system were diverse.} & 5.83 & 1.27 &  \\
& Comprehensive & \emph{The insight types used in the system are comprehensive.} & 4.50 & 1.62 &  \\
\midrule
\multirow{3}{*}{\textbf{Attribute types} (RQ2)} 
& Useful & \emph{The supported attribute types were useful.} & 5.92 & 1.16 &  \\
& Comprehensive & \emph{The supported attribute types were comprehensive.} & 5.25 & 1.71 &  \\
& Expressive & \emph{The attribute type combinations 
were useful.} & 5.58 & 1.38 &  \\
\midrule
\multirow{2}{*}{\textbf{Insight-type Ranking} (RQ3)}
& Useful & \emph{The ordering of insight types were useful.} &  3.92 & 1.83 &  \\
& Diversity & \emph{The top-ranked insight types were diverse.} & 4.58 & 1.73 & \\
\midrule
\multirow{2}{*}{\textbf{Insight Ranking} (RQ3)} 
& Useful & \emph{The insights were ranked appropriately within each row.} & 4.58 & 1.93 &  \\
& Diversity & \emph{The top-ranked insights in each row were diverse.} & 5.00 & 1.60 &  \\
\midrule
\multirow{4}{*}{\textbf{Visual Recommendations} (RQ4)} 
& Useful & \emph{The visualizations helped me understand the insights.} & 5.50 & 1.09 &  \\
& Diversity & \emph{The recommended visualizations were diverse.} & 5.50 & 1.38 & \\
& Aesthetic & \emph{The visualizations were aesthetically pleasing.} & 4.75 & 1.82 &  \\
& Easy to understand & \emph{The recommended visualizations were easy to understand.} & 4.42 & 1.08 &  \\
\midrule
\multirow{2}{*}{\textbf{User Interface} 
(RQ5)} 
& Useful & \emph{I found the UI features to be useful.} & 5.50 & 1.00 &  \\
& Expressive & \emph{It was useful to select the attributes of interest to filter insight recommendations.} & 6.50 & 0.80 &  \\
\bottomrule
\multicolumn{4}{l}{$^{*}$ 1=strongly disagree, 7=strongly agree.}
\end{tabular}
\vspace{-2mm}
\end{table}

\subsection{Participant Feedback} 
\label{sec:exp-user-study-feedback}
Overall feedback based on the Likert questions is provided in Table 4. In general, participants were positive about the system. When asked about what they liked most about the system, participants particularly appreciated that the system was comprehensive and automated when recommending insights. Even for participants who might be able to automate the process, they appreciated the ease-of-use, noting that \textit{“I do a lot of this by hand using Python so I definitely can see value in some immediate analysis. I end up generating similar graphs but each takes time (and learning in Python)”} (P3) or \textit{``I like that the visualizations are automated for me, so it’s great to just bookmark and view them''} (P1). Participants also noted that it would be particularly useful for exploratory data analysis ($mean=6.2$) and it helped them understand the data better ($mean=6.3$). P5 noted it was \textit{``a good way to start my orientation in a dataset''}. P3 noted how the system provided \textit{``immediate insights''} and could \textit{``save a lot of time exploring the dataset to understand it''}.

Participants were mixed about the comprehensiveness of the insight-types ($mean=4.5$). Most participants were able to find insights from the data based on the available insight-types, especially that they would not have otherwise found. P2 noted that the system included insight-types that they would not think to use, \textit{``I wouldn’t think to correlate the numerical values with each other off the top of my head, but that makes sense."}. P7 further explained, \textit{``a number of analyses listed were very useful in parsing out trends about the overall data.} Some expressed interest in insights related to specific slices of the data. For example, P2 suggested adding \textit{``comparisons between the categorical variables''} and P8 expressed a need for filtering by categorical values: \textit{``I don't know how I could select data from only variable (like city) and then see the results.''}

The visualizations were rated highly ($mean=5.5$), especially with the inclusion of annotations, which \textit{``help a lot to explain [the insights]''} (P2). Many participants commented about the diversity of chart types, citing this as a major strength of the system. \textit{``The system provide more advanced visualizations that other tools provide"} (P11) \textit{``Some of the graphs were very visually appealing and indicated trends across different categories very clearly. Definitely worth using this tool for those graphs alone.''} (P7)

The most common suggestion for improvement was to include more descriptive statistics and explanation of the insights used. Participants felt that the use of insight-type were too technical for people without significant statistics background. \textit{``Explanations about visualizations or statistical analysis would help me better explain what each chart is trying to tell me.''} (P3). Participants also suggested common improvements on the visualizations themselves, such as enforcing consistency between visualizations (e.g., color), providing interactions such as zooming, and details-on-demand to allow for analysis beyond EDA. These features in particular would help users in the next step of their investigative analysis, especially as they generate additional hypotheses, especially when comparing between multiple charts. Lastly, many participants (6/12) wanted more control over the visualizations (e.g., customizing the binning and configuring chart variables and axes).

\subsection{Discussion}
\label{sec:discussion}
\noindent\emph{Insight-Centric Visualization.}
The proposed visual insight recommendation framework offers greater generalizability as it applies to a significantly larger class of problems and which includes prior work as a very simple special case. 
Furthermore, our proposed system supports a greater granularity of insight types. Our study results indicate that the proposed class of visualization recommendation systems is successful in allowing users to discover insights quickly and gain a better understanding of the data.

\medskip\noindent\emph{Insight and Insight Type Ranking.}
By ranking insight-types and ranking insights within each type, we reduce the amount of time for users to find the most valuable insights while also allowing the users to explore the data according to the insight-types of interest.
Our results indicate that users generally found the ranked order of insights to be useful, and most of the insights users bookmarked were from the top ranked insights within a given insight-type. Top ranked insights were mostly found within the top 5 ranked insight types as well, and users generally agreed that the insight types themselves were useful and diverse.  

\medskip\noindent\emph{Limitations.}
There are a few key limitations to Spotlight. 
Hypothesis driven data exploration could be further improved by the addition of several other small enhancements to the system, such as enabling filtering by values or ranges within a variable, allowing users to customize chart axes, and enabling visualization interactivity. 
A limitation of the user study we conducted was the use of datasets with a small 
number of columns (less than 10).
While users still found the insight and insight-type recommendations useful, 
it would have been more interesting to investigate the utility of the insight-type and insight rankings using datasets with significantly more columns (\eg, 50-100). 
In this case, appropriately ranking insight-types and insights within each type become significantly more important to the user as the number of possible visual insights increases exponentially with the amount of attributes in the dataset.

\section{Conclusion} \label{sec:conc}
In this work, we proposed a visual insight recommendation system that makes it easy and fast for users to find interesting and important insights in their data visually.
Given an arbitrary dataset, we first recommend the top-k insights and their visualizations for each insight-type (\eg, time-series outliers).
The visual insights for each insight-type are then ranked and displayed to the user in a single row.
More importantly, to enable the user to find interesting insights fast, we also score and rank the insight-types where each insight-type corresponds to a row of visual insights.
By recommending to the user the top-r insight-types (rows of visual insights), this makes it easy for the user to identify important and relevant insights in their data quickly.
Finally, future work will investigate personalizing the insight-type ranking and ranking of insights within each insight-type based on user behavior logs (\eg, bookmarked insights, and insight-types of those bookmarked insights, insights a user clicked on and so on).

\bibliographystyle{abbrv}
\bibliography{paper} 
\end{document}